\documentclass[12pt,preprint]{aastex}

\shorttitle{AKARI NIR spectroscopy of YSOs in the LMC}
\shortauthors{Shimonishi et al.}

\begin{document}

\title{$\textit{AKARI}$ Near Infrared Spectroscopy: Detection of H$_2$O and CO$_2$ Ices toward Young Stellar Objects in the Large Magellanic Cloud}

\author{Takashi SHIMONISHI\altaffilmark{1,2}, Takashi ONAKA\altaffilmark{1}, Daisuke KATO\altaffilmark{1}, Itsuki SAKON\altaffilmark{1}, \\
Yoshifusa ITA\altaffilmark{3}, Akiko KAWAMURA\altaffilmark{4}, and Hidehiro KANEDA\altaffilmark{5}}

\altaffiltext{1}{Department of Astronomy, Graduate School of Science, The University of Tokyo, 7-3-1 Hongo, Bunkyo-ku, Tokyo 113-0003, Japan}
\altaffiltext{2}{shimonishi@astron.s.u-tokyo.ac.jp}
\altaffiltext{3}{National Astronomical Observatory of Japan, 2-21-1 Osawa, Mitaka, Tokyo 181-8588, Japan}
\altaffiltext{4}{Department of Astrophysics, Nagoya University, Chikusa-ku, Nagoya 464-8602, Japan}
\altaffiltext{5}{Institute of Space and Astronautical Science, Japan Aerospace Exploration Agency, 3-1-1 Yoshinodai, Sagamihara, Kanagawa 229-8510, Japan}

\begin{abstract}
 We present the first results of $\textit{AKARI}$ Infrared Camera near-infrared spectroscopic survey of the Large Magellanic Cloud (LMC). 
We detected absorption features of the H$_2$O ice 3.05$\mu$m and the CO$_2$ ice 4.27$\mu$m stretching mode toward seven massive young stellar objects (YSOs). 
These samples are for the first time spectroscopically confirmed to be YSOs. 
We used a curve-of-growth method to evaluate the column densities of the ices and derived the CO$_2$/H$_2$O ratio to be 0.45$\pm$0.17. 
This is clearly higher than that seen in Galactic massive YSOs (0.17$\pm$0.03). 
We suggest that the strong ultraviolet radiation field and/or the high dust temperature in the LMC may be responsible for the observed high CO$_2$ ice abundance. 
\end{abstract}


\section{Introduction}
 Properties of extragalactic young stellar objects (YSOs) provide us important information on the understanding of the diversity of YSOs in different galactic environments. The Large Magellanic Cloud (LMC), the nearest irregular galaxy to our Galaxy \citep[$\sim$50kpc;][]{Alv04}, offers an ideal environment for this study since it holds a unique metal-poor environment \citep{Luc98}. Because of its proximity and nearly face-on geometry, various types of surveys have been performed toward the LMC \citep[e.g., ][ and references therein]{Zar04,Mei06,Kat07}.

 An infrared spectrum of YSOs shows absorption features of various ices which are thought to be an important reservoir of heavy elements and complex molecules in a cold environment such as a dense molecular cloud or an envelope of a YSO \citep[e.g., ][]{Chi98,Num01,Whit07,Boo08}. 
These ices are thought to be taken into planets and comets as a result of subsequent planetary formation activity \citep{Ehr00}. 
Studying the compositions of ices as functions of physical environments is crucial to understand the chemical evolution in circumstellar environments of YSOs and is a key topic of astrophysics. 
H$_2$O and CO$_2$ ices are ubiquitous and are major components of interstellar ices \citep{vDB98,BE04}. 
Since the absorption profile of the ices is sensitive to a chemical composition of icy grain mantles and a thermal history of local environments, the ices are important tracers to investigate the properties of YSOs. 
However, our knowledge about the ices around extragalactic YSOs is limited because few observations have been performed toward extragalactic YSOs. 
Therefore infrared spectroscopic observations toward YSOs in the LMC are important if we are to improve our understanding of the influence of galactic environments on the properties of YSOs and ices.

 $\textit{AKARI}$ is the first Japanese satellite dedicated to an infrared astronomy launched in February 2006 \citep{Mur07}. We have performed a near infrared spectroscopic survey of the LMC using a powerful spectroscopic survey capability of Infrared Camera \citep[IRC;][]{TON07} on board $\textit{AKARI}$. 
In this letter, we present 2.5--5$\mu$m spectra of newly confirmed YSOs in the LMC with our survey, and discuss the abundance of H$_2$O and CO$_2$ ice.

\section{Observations and Data Reduction}
 The observations reported here were obtained as a part of the $\textit{AKARI}$ IRC survey of the LMC \citep{Ita08}. 
An unbiased slitless prism spectroscopic survey of the LMC has been performed since May 2006. 
In this survey, the IRC02b $\textit{AKARI}$ astronomical observing template (AOT) with the NP spectroscopy mode was used to obtain low resolution spectra ($\textit{R}$ $\sim$ 20) between 2.5 and 5 $\mu$m. 
By June 2008, 621 pointing observations were performed toward the LMC, and about 10 deg$^2$ area was observed. 
The near-infrared (NIR) spectroscopic survey is continuing even after the exhaustion of the liquid helium in August 2008. 
 
 The spectral analysis was performed using the standard IDL package prepared for the reduction of $\textit{AKARI}$ IRC spectra \citep{Ohy07}. 
Raw data were converted to dark-subtracted, linearity-corrected, and flat-field-corrected frames. 
Three pixels were integrated in the spatial direction to extract a point-source spectrum from the slitless spectroscopic image. 
Background levels of each source were estimated from the signal counts of the adjacent regions on both sides of the target spectrum. 
The wavelength calibration accuracy is estimated to be about $\sim$0.01$\mu$m \citep{Ohy07}

\section{The selection of YSOs}
 We select infrared-bright objects from the point-source catalog of the $\textit{Spitzer}$ SAGE project \citep{Mei06} with the following selection criteria: 
 (1)[3.6] -- [4.5] $\textgreater$ 0.3 and [5.8] -- [8.0] $\textgreater$ 0.6, and (2)[3.6] $\textless$ 12 and [4.5] $\textless$ 11.5, 
 where [wavelength] represents the photometric value in magnitude at each wavelength in microns. 
Criterion (1) refers to the YSO model of \citet{Whin04}, and criterion (2) comes from the detection limit of the $\textit{AKARI}$ IRC NP mode. 
This rough selection is applied to the sources located in the survey area of $\textit{AKARI}$ IRC, and about 300 sources are selected. 
These photometrically selected sources include not only massive YSOs, but also a large number of dusty evolved stars since their infrared spectral energy distributions (SEDs) are similar to each other. 
For the accurate selection of YSOs, we select the sources that show absorption features of the 3.05 $\mu$m H$_2$O ice and the 4.27 $\mu$m CO$_2$ ice stretching mode in their NIR spectra taken by the present spectroscopic survey. 
The presence of CO$_2$ ice is strong evidence of YSOs since the detection of CO$_2$ ice toward dusty evolved stars has not been reported \citep{Bar99,Syl99}. 
Spectral overlapping with other sources located in a dispersion direction is a serious problem for slit-less spectroscopy, which makes it difficult to obtain reliable spectra. 
We check the overlapping contamination by visual inspection, and we only use the sources without such contamination in the following analysis. 

 As a result, we spectroscopically confirmed seven massive YSOs in the LMC for the first time. 
The sources are listed in Table 1 with the observation parameters.
Six of the seven sources are included in the recent YSO candidates catalog \citep{Whin08}, and one source (ST6) is a newly found YSO. 
The spectra of these sources are shown in Fig 1 together with the results of spectral fitting (see $\S$4 for details). 
The absorption features of H$_2$O and CO$_2$ ices are rather broadened due to the low spectral resolution of the $\textit{AKARI}$ IRC NP spectroscopy mode but clearly detected. 
This is the first clear detection of the 4.27 $\mu$m stretching mode of CO$_2$ ice toward extragalactic YSOs. 
In addition, unresolved emission of PAHs and the hydrogen recombination line Pf$\delta$ around 3.3 $\mu$m, the Br$\alpha$ line at 4.05 $\mu$m and blended absorption features of 4.62 $\mu$m XCN and 4.67 $\mu$m CO ices around 4.65 $\mu$m are detected toward several sources. 
However, it is difficult to evaluate the column densities of XCN and CO ices accurately with the present low spectral resolution data.

\section{Spectral Fitting}
 We fit a polynomial of the second to fourth order to the continuum regions and divide the spectra by the fitted continuum (Fig 1). 
The wavelength regions for the continuum are set to be 2.5--2.7 $\mu$m, 3.6--3.7 $\mu$m, 4.0--4.15 $\mu$m, and 4.9--5.0 $\mu$m \citep{Gib04}. 

 Due to the low spectral resolution of the IRC NP spectroscopy mode, direct comparison of the observed spectra with the absorption profiles of laboratory ices is difficult. 
However, the equivalent width of absorption does not depend on the spectral resolution of the spectrum. 
Therefore we used a curve-of-growth method to derive the column densities of the ices. 
A Gaussian profile with the fixed central wavelength is fitted to the absorption bands to derive the equivalent width (Fig 1). 
The vertical axis of the plot is shown in units of the normalized flux because it is difficult to estimate the optical depth directly from the present low resolution data. 
We use the laboratory absorption profiles of H$_2$O and CO$_2$ ices taken from the Leiden Molecular Astrophysics database \citep{Ehr96} to calculate the curve of growth. 
The profiles of pure H$_2$O ice and mixture of H$_2$O : CO$_2$ (100:14) ice both at 10K are used for the calculation since these compositions are typical in the interstellar ices \citep{Num01,Gib04}.
The present spectrum cannot resolve the polar and apolar CO$_2$ ice features, and the present analysis assumes the polar CO$_2$ ice only. 
However, contribution of the apolar ice is generally small toward YSOs \citep{Ger95,Ger99}. 
The FWHM of the absorption profiles of H$_2$O and CO$_2$ ices change by approximately 10 $\%$ depending on the compositions of the ices \citep{Ger99,Gib04}. 
We confirm that this range of the variation in the FWHM makes negligible effects on the derived column densities compared with the observational errors. 
We adopt the band strengths of H$_2$O and CO$_2$ ices to be 2.0$\times$10$^{-16}$ and 7.6$\times$10$^{-17}$ cm molecule$^{-1}$ \citep{Ger95}, respectively. 
The derived column densities are listed in Table 1.

 Since this study applies the curve of growth method to the calculation of the ice column density for the first time, we check the validity of our method. 
We use a few Galactic YSO spectra taken by $\textit{ISO}$ SWS whose column density of H$_2$O and CO$_2$ ices are derived in \citet{Gib04}. 
These high-resolution spectra are converted to low-resolution spectra by convolving the slit function of the NP spectroscopy mode, and then we derive the column density for these converted spectra using the same method described above. 
The comparison of the obtained column density with the value presented in \citet{Gib04} shows that the differences are within 10 $\%$ and 20 $\%$ for the H$_2$O ice and the CO$_2$ ice, respectively. 
The result plotted in Fig 2 includes this uncertainty in addition to the observational error.

\section{Results and Discussion}
 The obtained column densities of H$_2$O and CO$_2$ ices are plotted in Fig 2. 
The error bars become larger for the larger column density due to the saturation effect of the curve-of-growth. 
A linear fit to the data points indicates that the CO$_2$/H$_2$O ice column density ratio in the LMC is 0.45 $\pm$ 0.17. 
The large uncertainty mainly comes from the errors in the curve-of-growth analysis. 
For comparison, column densities of Galactic massive YSOs taken from \citet{Gib04} and their CO$_2$/H$_2$O ice column density ratio of 0.17 $\pm$ 0.03 \citep{Ger99} are also plotted in Fig 2. 
A similar CO$_2$/H$_2$O ratio of 0.18 $\pm$ 0.04 is also observed toward a Galactic quiescent dark cloud \citep{Whit07}, 
while a relatively high CO$_2$/H$_2$O ratio of 0.32 $\pm$ 0.02 is observed toward Galactic low-- and intermediate-- mass YSOs, and some of them reach $\sim$0.4 \citep{Pon08}. 
Although the uncertainty is large, it is clear from the present results that the CO$_2$/H$_2$O ice ratio in the LMC is higher than the typical ratios of the Galactic objects. 
Since the distribution range of the H$_2$O ice column density in the LMC is comparable to that of the massive Galactic YSOs, it can be concluded that the abundance of the CO$_2$ ice is higher in the LMC. 
The present results suggest that the different galactic environment of the LMC is responsible for the high CO$_2$ abundance.

The formation mechanism of CO$_2$ ice in circumstellar environments of YSOs is not understood, however a number of scenarios have been proposed. 
Several laboratory experiments indicate that the CO$_2$ ice is efficiently produced by UV photon irradiation to H$_2$O-CO binary ice mixtures \citep[e.g., ][]{Wat07}. 
The LMC has an order-of-magnitude stronger UV radiation field than our Galaxy due to its active massive star formation \citep{Isr86}, 
which could lead to the higher CO$_2$/H$_2$O ratio in the LMC. 
The high CO$_2$/H$_2$O ratio toward a YSO in the LMC is also reported in \citet{vanL05}, and they suggest that a different radiation environment in the LMC is one of the reasons for the high CO$_2$ abundance. 
On the other hand, the model of diffusive surface chemistry suggests that high abundance of CO$_2$ ice can be produced at relatively high dust temperatures \citep{Ber99,Ruf01} . 
Several studies have reported that the dust temperature in the LMC is generally higher than in our Galaxy based on far-infrared to submillimeter observations of diffuse emission \citep[e.g., ][]{Agu03, Sak06}. 
Therefore the high dust temperature may also have an effect on the high CO$_2$ ice abundance in the LMC.

 It is difficult to separate the effect of the UV radiation field and the dust temperature on the high abundance of CO$_2$ ice in the LMC by our low-resolution NIR spectra. 
The 4.62 $\mu$m XCN feature is known to be indicative of strong UV irradiation \citep{Ber00,Spo03}. 
On the other hand, detailed profile analysis of the 3.05 $\mu$m H$_2$O ice stretching mode and the 15.2 $\mu$m CO$_2$ ice bending mode should reveal the temperature and compositions of the ices \citep{Ehr96,Obe07}. 
Future observations of the XCN feature and the H$_2$O and CO$_2$ ice features with a sufficient wavelength resolution will be useful to investigate this problem.

\section{Summary}
 We performed a near-infrared spectroscopic survey of the LMC with $\textit{AKARI}$ IRC. 
We spectroscopically confirmed seven massive YSOs that show absorption features of H$_2$O and CO$_2$ ices. 
This is the first detection of the 4.27 $\mu$m CO$_2$ ice feature toward extragalactic YSOs. 
The derived ice column densities indicate that the abundance of CO$_2$ ice is clearly higher in the LMC than our Galaxy. 
The relatively strong UV radiation field and/or high dust temperature in the LMC may be responsible for the observed high abundance of CO$_2$ ice. 
Our study shows the difference in the chemical composition around extragalactic YSOs, suggesting that extragalactic YSOs hold quite different environments from Galactic ones.

\acknowledgments	
 {\it AKARI} is a JAXA project with the participation of ESA. 
We thank all the members of the {\it AKARI} project for their continuous help and support. 
We also thank Youichi Ohyama for great support in the analysis of spectroscopic data, and we thank the $\textit{AKARI}$ IRC team and the LMC team for their continuous support. 
This work is supported by a Grant-in-Aid for Scientific Research from the JSPS.

\clearpage

\begin{deluxetable}{ccccccccc}
\tabletypesize{\scriptsize}
\tablecaption{OBSERVATION PARAMETERS AND COLUMN DENSITIES OF ICES  \label{tbl1}}
\rotate
\tablewidth{0pt}
\tablehead{
\colhead{Number} & \colhead{AKARI ID} & \colhead{Obs. ID} & \colhead{Obs. Date} & \colhead{Other Name} & \colhead{RA[J2000]} & \colhead{DEC[J2000]} & \colhead{N(H$_2$O)} & \colhead{N(CO$_2$)\tablenotemark{a}}  \\
\colhead{} & \colhead{} & \colhead{} & \colhead{} & \colhead{} & \colhead{} & \colhead{} & \colhead{(10$^{17}$ cm$^{-2}$)} & \colhead{(10$^{17}$ cm$^{-2}$)}
}
\startdata
ST1 & J053931-701216 & 2211375 & 2007 Apr 12 & 05393117-7012166\tablenotemark{a} & 05:39:31.15 & -70:12:16.8 & 9.6$_{-1.9}^{+1.9}$    &  6.7$_{-3.6}^{+4.8}$    \\
ST2 & J052212-675832 & 2210229 & 2006 Nov 24 & NGC 1936                          & 05:22:12.56 & -67:58:32.2 & 11.1$_{-1.4}^{+1.6}$   &  3.1$_{-1.5}^{+1.7}$    \\
ST3 & J052546-661411 & 2213132 & 2007 Jun 22 & ......\tablenotemark{b}           & 05:25:46.69 & -66:14:11.3 & 29.7$_{-4.6}^{+4.5}$   &  15.5$_{-9.4}^{+13.8}$ \\
ST4 & J051449-671221 & 2200073 & 2006 Jun 6  & IRAS F05148-6715                  & 05:14:49.41 & -67:12:21.5 & 18.7$_{-2.3}^{+2.3}$   &  8.2$_{-3.7}^{+4.5}$    \\
ST5 & J053054-683428 & 2213061 & 2007 Mar 13 & IRAS 05311-6836                   & 05:30:54.27 & -68:34:28.2 & 31.7$_{-4.5}^{+4.7}$   &  12.4$_{-6.8}^{+9.1}$   \\
ST6 & J053941-692916 & 2201137 & 2006 Oct 25 & 05394112-6929166\tablenotemark{a} & 05:39:41.08 & -69:29:16.8 & 59.1$_{-25.0}^{+38.4}$ &  ...                    \\
ST7 & J052351-680712 & 2210220 & 2006 Nov 29 & IRAS05240-6809                    & 05:23:51.15 & -68:07:12.2 & ...                    &  55.3$_{-30.3}^{+40.3}$ \\
\enddata
\tablenotetext{a}{2MASS ID}
\tablenotetext{b}{The source is in a cluster}

\end{deluxetable}

\clearpage

\begin{figure}
\figurenum{1}
\epsscale{0.45}
\plotone{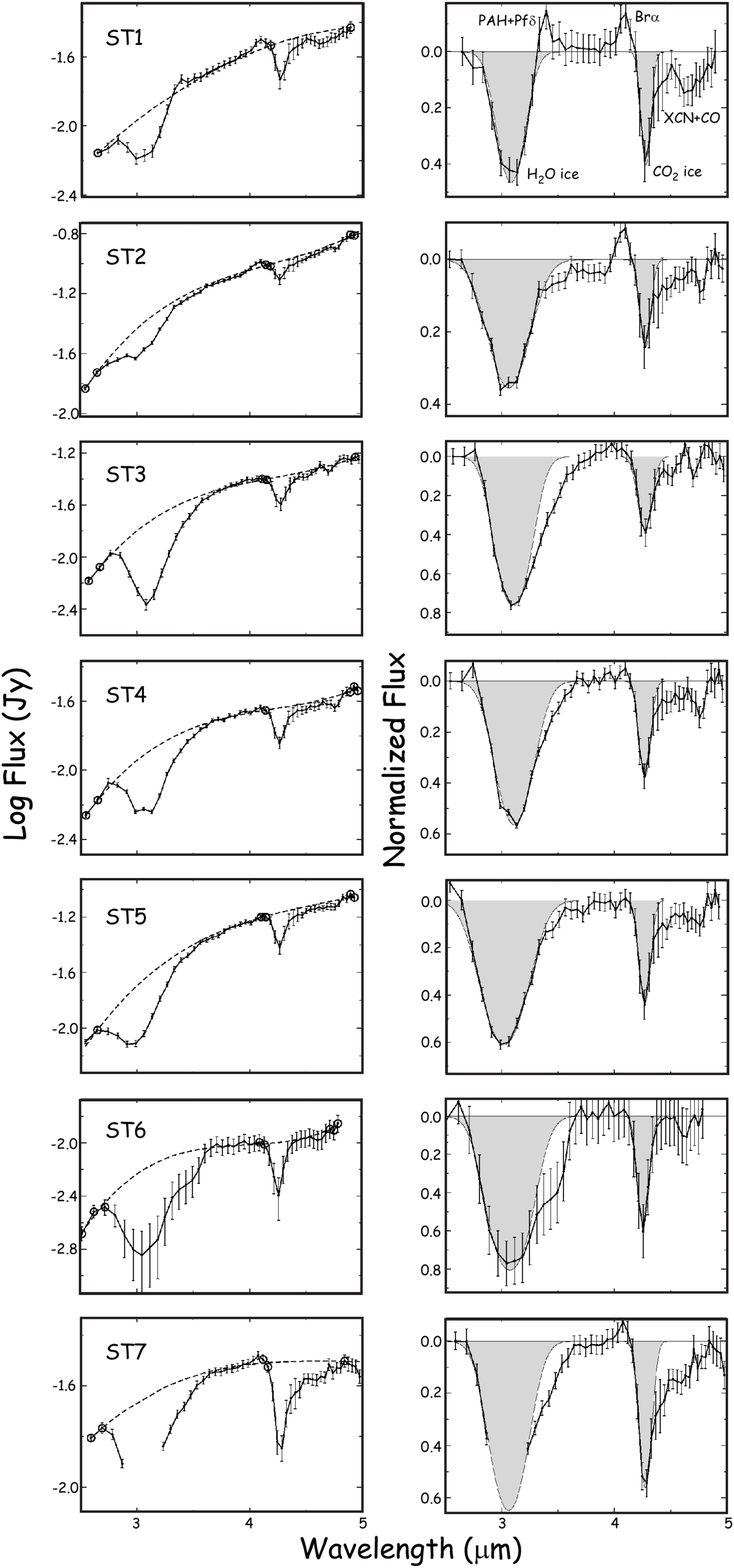}
\caption{{\small $\textit{AKARI}$ IRC 2.5--5$\mu$m spectra of YSOs in the LMC. 
Left: Plots of log$_{10}$ flux(Jy) vs. wavelength($\mu$m). 
Open circles represent the points that used for the continuum determination. Dashed lines represent derived continuum.
Right: Plots of normalized flux (1 -- F/F$_c$; F and F$_c$ represent observed flux and continuum flux, respectively) vs. wavelength($\mu$m). 
Shaded regions around 3.0$\mu$m and 4.3$\mu$m represent fitted Gaussian for absorption features of H$_2$O and CO$_2$ ices, and the areas correspond to the equivalent width.
The positions of H$_2$O, CO$_2$, XCN, CO ice absorption bands, PAH emission bands, and hydrogen recombination lines are shown.}
\label{fig1}}
\end{figure}

\clearpage

\begin{figure}
\figurenum{2}
\epsscale{0.9}
\plotone{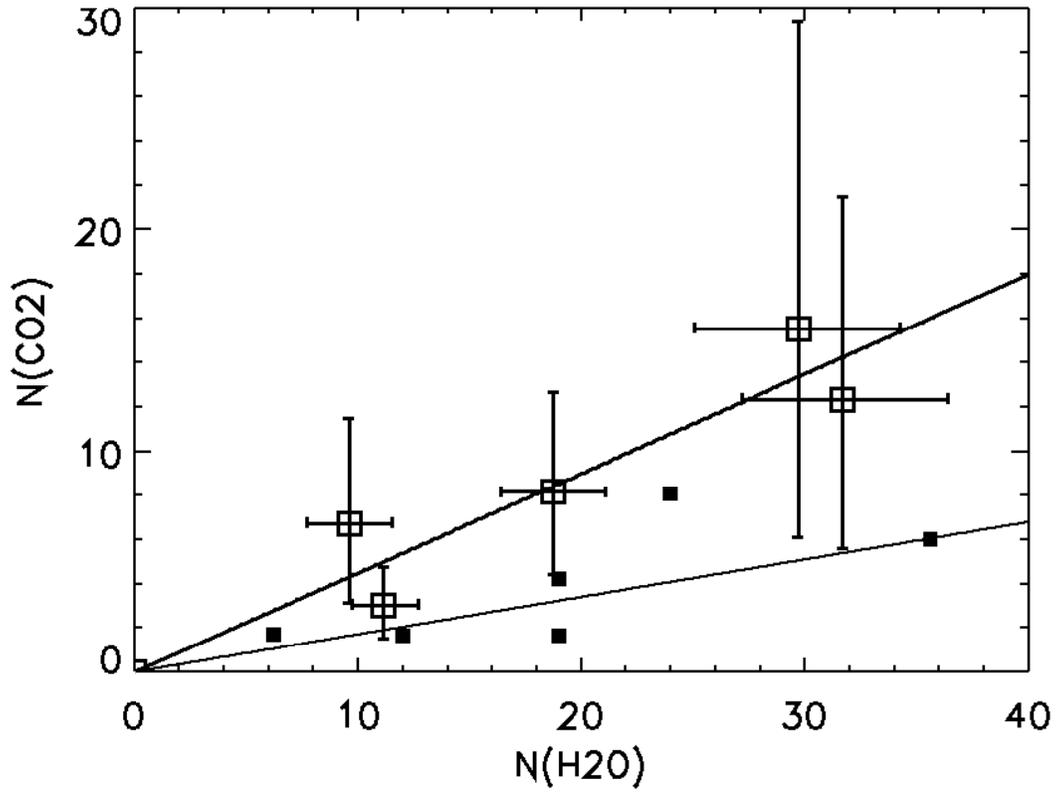}
\caption{{\small CO$_2$ ice vs. H$_2$O ice column density in a unit of 10$^{17}$ cm$^{-2}$. Open squares with error bars represent the results of this study. Filled squares represent those of Galactic massive YSOs \citep{Gib04}. Upper and lower solid lines represent CO$_2$/H$_2$O $\sim$ 0.45 and 0.17, respectively. The sources ST6 and ST7 are not plotted due to their large errors.}
\label{fig2}}
\end{figure}

\end{document}